\documentstyle[aps,prl,preprint]{revtex} 

\begin{document}

\draft

\title{The Trouble with  Quantum Bit Commitment}
\author{Dominic Mayers}
\address{D\'epartement IRO, Universit\'e de Montr\'eal \\
C.P. 6128, succursale Centre-Ville,Montr\'eal (Qu\'ebec), Canada H3C 3J7.}
\date{\today}
\maketitle

\begin{abstract}
In a recent paper, Lo and Chau explain how to break a family of
quantum bit commitment schemes, and they claim that their attack applies
to the 1993 protocol of Brassard, Cr\'epeau, Jozsa and Langlois (BCJL).
The intuition behind their attack is correct, and indeed they expose a
weakness common to all proposals of a certain kind, but the BCJL protocol
does not fall in this category.  Nevertheless, it is true that the BCJL
protocol is insecure, but the required attack and proof are more subtle.  
Here we provide the first complete proof that the BCJL protocol is insecure.
\end{abstract}

\paragraph{Introduction}     
Recently, Lo and Chau have made available on the {\tt quant-ph}
archives~\cite{lc96a} a preprint that explains how to break a family of
quantum bit commitment schemes, and they claim that their attack applies to
the protocol of Brassard, Cr\'epeau, Jozsa and Langlois~\cite{bcjl93}, 
hereafter
called the BCJL protocol.   The~intuition behind
their attack against the BCJL protocol is correct, and indeed they expose a
weakness common to all proposals of a certain kind
(including~\cite{ardehali95,bb84}), but the BCJL protocol
does {\em not\/} fall in this category (see~the opening paragraph in
Section~\ref{simplecase}).  Nevertheless, it is true that the BCJL protocol
is insecure, but the proof which we have known for quite some
time~\cite{mayers95b}, is more subtle.  Here we provide this first complete
proof that the BCJL protocol is insecure.

We have also considered several variations on the BCJL theme.
Neither Lo and Chau's attack nor the correct attack on the
BCJL protocol explained below apply to these variations.
One~of these variations consists in having the photons travel
in the reverse direction compared with the original BCJL protocol.
This~is natural for many cryptographic applications.
Nevertheless, all these variations fail as well for different reasons related 
to subtle points in quantum information theory~\cite{hughston93,jozsa94a}
that only began to be understood at the time the BCJL paper was written.  
A~proof that none of these variations work will be the subject
of a forthcoming paper: the~current paper focuses on the correct attack
against the original BCJL protocol.  

Lo and Chau wrote: ``The security of other quantum cryptographic
protocols say for oblivious transfer [...] remains to be examined.''
This is a serious concern because quantum
oblivious transfer and many other quantum protocols
depend on the security of bit commitment~\cite{bbcs92,crepeau94,yao95}. 
On~the other hand, we disagree with the following sentence from Lo and Chau:
``One~might wonder if all of quantum cryptography may stumble under closer
scrutinies'' because our earlier proof of security for quantum
key distribution~\cite{mayers95a,mayers96c} would hold even if secure quantum
bit commitment is not possible despite the fact that it is
based on an earlier ``proof'' of security for quantum oblivious transfer
that fails in the absence of a secure bit commitment scheme.
The reason is that the proof of security for quantum key distribution
does not depend on the security of quantum oblivious transfer, but
rather on the (correct) proof that quantum oblivious transfer would
be secure if implemented on top of a secure bit commitment scheme.
     
\paragraph{Bit Commitment}     
Any cryptographic task defines the relationship between      
inputs and outputs respectively     
entered and received by the task's participants.       
In bit commitment, Alice enters a bit $b$.     
At a later time,  Bob may request this bit and,     
whenever he does, he receives this bit,      
otherwise he learns nothing about $b$.      
     
In a naive but concrete realization of bit commitment,     
Alice puts the bit into a strong-box     
of which she keeps the key and then     
gives this strong-box to Bob.  At a later time,     
if Bob requests the bit, Alice gives the key to Bob.       
The main point is that Alice cannot change her mind     
about the bit $b$ and Bob learns nothing about it unless he     
obtains the key.       
Now, let us sketch the BCJL protocol.
\begin{enumerate}
\item[]\hspace{-\leftmargin}COMMIT(b)
\item Bob chooses a linear code $C$ (with some required properties) and     
announces it to Alice.
\item Alice chooses a perfectly random string $r \in \{0,1\}^n$     
and announces it to Bob.
\item Alice chooses a perfectly random string $\theta \in \{+,\times\}^n$
and a string $c$ uniformly distributed over  
$\{c \in C \; | \; c \odot r = b\}$.     
\item Alice sends $n$ photons to Bob in a product state      
$|c\rangle_{\theta} = |c_1 \rangle_{\theta_1} \otimes      
\ldots \otimes |c_n \rangle_{\theta_n}$.      
\item Bob measures the $n$ photons in a perfectly random basis     
$\hat{\theta} \in \{+, \times \}^n$     
and obtains the classical outcome $\hat{c} = \hat{c}_1 \ldots \hat{c}_n$.     
\end{enumerate}     
To unveil the bit $b$, Alice announces $\theta$, $c$ and $b$     
to Bob.  Bob computes a function      
$T_b(\theta, c, \hat{\theta}, \hat{c},C,r)$ to test     
whether or not he should accept Alice's claim.  The function
$T_b$ returns ${\rm ok}$ if if the bit $b$ announced by Alice
is accepted, otherewise $T_b$ returns ${\rm not~ok}$.  The exact     
description of the function $T_b$ is irrelevant for     
our analysis.  One does not need to understand in detail
how the BCJL protocol attempts to realize bit commitment to see
that it cannot work.  
The pair $(C,r)$ corresponds to the information that is shared     
between Alice and Bob just before Alice sends the     
photons.  Most of our analysis is done for $(C,r)$ fixed,     
so $(C,r)$ is suppressed in most of the notations.      
For instance, from here on, $(C,r)$ is suppressed in the input of     
the functions $T_b(\theta,c,\hat{\theta}, \hat{c})$.     

We denote $p(\theta,c|b)$ the probability of     
the state $|c \rangle_{\theta}$ given the bit $b$ that Alice has in mind.    
Such a random distribution of pure states is called a mixture.
The~BCJL authors~\cite{bcjl93} explains that, given $b$,      
the results of any physical measurement whatever on the      
mixture prepared by Alice depend only on the density operator
$\rho_b = \sum_{\theta,c} p(\theta,c|b)\; (|c\rangle\langle c|)_{\theta}$.
This means that a dishonest Alice could      
send any other mixture $\{(|\psi\rangle, p(\psi))\}$ such      
that $\sum_\psi p(\psi) |\psi\rangle \langle \psi| = \rho_b$,
for some $b \in \{0, 1\}$, without being detected.
It~is correctly shown in~\cite{bcjl93}
that such a strategy does not work.  Their Theorem~3.7   
implies that, for all practical purposes,      
any pure state $|\psi\rangle$ commits      
Alice to a single bit~$b$.      
More precisely, once Bob has received     
the $n$ photons in a pure state $|\psi\rangle$,     
there exists one value $b$ such that,      
except with a negligible probability, Alice cannot convince Bob     
that she had the opposite bit $\bar{b}$ in mind, that is,      
$(\forall \; \theta, c)$,      
$\Pr(T_{\bar{b}}(\theta, c, \hat{\theta}, \hat{c}) = ok\; 
|\; \Psi = \psi)$     
is exponentially small.  The conclusion in~\cite{bcjl93} is that      
the protocol is secure against Alice.     
     
However, as explained by these authors,     
preparing a mixture is not the only way to prepare a density     
matrix.  Alice may prepare the density matrix $\rho_b$ of      
the $n$ photons by introducing another system $A$      
kept on her side and preparing the new incremented system      
in a pure state $|\phi\rangle$.  Note that a pure state      
$|\phi\rangle$ of the incremented system is  not  in general     
the product of a pure state of $A$ with a pure state of the $n$ photons.     
 
This possibility was considered in~\cite{bcjl93}.     
In~the appendix of their paper, they mention the true fact     
that as far as the results of Bob's measurement is concerned,      
this alternative approach is equivalent to a preparation of      
a mixture by Alice.       
It is true that no matter what Alice does on her side,     
at the best, she will be found in a situation where      
everything behaves as if such a mixture had been sent to Bob.     
For any such a mixture, it is true that Alice may only open     
a single bit $b$.  However, this is not sufficient to show the     
security against Alice.  The problem, which we explain in     
this paper, is that by delaying her measurement on the
system $A$ that she kept on her side Alice
may choose the mixture and thus the bit $b$ after the      
commit.

Let $W$ be the input     
$(\theta,c,\hat{\theta},\hat{c})$  of the functions    
$T_0$ and $T_1$.     
We consider $W$ as a random variable.  We denote $W^{(0)}$     
the random variable $W$ conditioned by $b = 0$      
and $W^{(1)}$ the same random variable conditioned      
by $b = 1$.  Both $W^{(0)}$ and $W^{(1)}$ refer to      
the honest protocol.  If the protocol is     
correct, except with negligible probability, we     
should have $T_b(W^{(b)}) = {\rm ok}$, that is, if     
Alice has been honest and has chosen the bit $b$,     
then Bob should accept it.  The problem with     
the protocol COMMIT is that a dishonest 
Alice, by delaying her measurement,
can choose after the commit to have $W$ behave either as $W^{(0)}$
or $W^{(1)}$.  Let $\psi$ be the classical outcome of this measurement.      
A dishonest Alice computes $(\theta,c)$ in view of $\psi$, and  announces     
$(b,\theta,c)$ to Bob.   In the following, without loss of generality,     
we may assume that $(\theta,c)$ is the classical outcome     
of a measurement ${\bf M}^{(A)}$ executed by Alice
because the  computation of $(\theta,c)$ may be considered     
as a part of this measurement.  
      
Alice can cheat if 
she can create a state $|\phi\rangle$ of the incremented     
system such that, for every $b$, there exists a measurement 
${\bf M}_b^{(A)}$  on the system $A$ such that
\begin{itemize}     
\item the classical outcome $(\theta,c)$ of ${\bf M}_b^{(A)}$
has the same probability distribution as the corresponding
pair $(\theta,c)$ in the honest case when Alice chooses $b$,     
\item whenever Alice obtains $(\theta,c)$, the $n$ photons on     
Bob's side collapse in the state $|c\rangle_{\theta}$ as in     
the honest protocol.     
\end{itemize}     
A dishonest Alice executes ${\bf M}_b^{(A)}$ after that Bob has executed     
his measurement on the $n$ photons.  However, these two     
measurements commute, that is, the random variable $W$     
is the same whether Alice measures before or after Bob,
and the only thing that matters is the distribution of $W$.      
Therefore, we may assume that Alice measures before Bob.       
In~such a case, the above condition says that after Alice's 
measurement the situation is     
exactly as in the honest protocol. Therefore, when      
the measurement ${\bf M}_b^{(A)}$ 
is chosen by Alice, we have $W = W^{(b)}$,     
that is, $W$ behaves as it would in the honest protocol when Alice     
chooses the bit $b$, and 
$T_b$ is expected to return $ok$.  So Alice can cheat.     
In the remainder of this paper, we show that     
for all practical purposes, if the protocol is
secure against Bob, then the above condition  holds.         
     
\paragraph{A simpler case}\label{simplecase}     
This section considers the security of any bit commitment
protocol in which Alice commits herself to $b$ by sending 
photons to Bob where the density matrices for \mbox{$b = 0$}
and \mbox{$b = 1$} are identical.  
This~is precisely the case that was independently considered by Lo and
Chau~\cite{lc96a}.
This~is sufficient to break the old protocol in~\cite{bb84}
(which is not surprising since a simple \mbox{EPR-type} attack
was already included in the same paper~\cite{bb84})
as well as a more recent protocol proposed by Ardehali~\cite{ardehali95}.
However, this analysis is insufficient to break the BCJL protocol
since the density matrices $\rho_b$ prepared by Alice, when she     
has respectively \mbox{$b =0$} and \mbox{$b =1$} in mind, are not identical. 
    
In~fact, the main thrust of the BCJL paper was to prove that
Bob could not cheat {\em despite\/} the fact that he was sent
slightly different density matrices by Alice depending on which bit she
wanted to commit~to.  (Clearly, the protocol could not be secure
if the density matrices had differed too much, because then
Bob would be able to distinguish between      
\mbox{$b=0$} and \mbox{$b=1$} without any     
help from Alice.)  In this section, we show that the above
mentioned condition  holds under the     
simplifying assumption \mbox{$\rho_0 = \rho_1$}. 
In~the next section, we shall consider the situation
that really applies to the BCJL protocol.     
       
In the commit part,  Alice prepares a pure state of the incremented     
system such that the density matrix for the $n$ photons is     
$\rho = \rho_0 = \rho_1$, that is,        
the same density matrix that would have been honestly     
prepared by Alice no matter whether she had $b =0$ or $b =1$ in mind.     
Now, let us assume     
that, after the commit, Alice wants to convince Bob that she had      
some bit $b$ of her choice in mind. It is shown in  \cite{hughston93}      
that by choosing the appropriate measurement on $A$,      
Alice may choose any mixture      
$\{(|\psi\rangle, p(\psi)\}$ such that      
$\sum_c p(\psi)      
|\psi \rangle \langle \psi| = \rho$.       
It is explained in \cite{hughston93} that when Alice chooses the mixture     
$\{(|\psi\rangle, p(\psi)\}$,  she receives      
the classical outcome $\psi$      
with probability $p(\psi)$ and, furthermore,  the 
classical information $\psi$ received by Alice  
uniquely determines the collapsed state $|\psi\rangle$ of the      
$n$ photons on Bob's side. 
In particular, Alice may choose $\{(|\psi\rangle, p(\psi)\}$ to be  
the mixture $\{(|c\rangle_{\theta}, p(\theta,c|b) )\}$. 
We have that the classical information $\psi$ received 
by Alice, which uniquely determines the collapsed state  
$|\psi\rangle = |c\rangle_{\theta}$ on Bob's side,  is $(\theta,c )$.   
This shows that the above mentioned condition holds.     
      
\paragraph{The real situation}     
Now, we consider the real situation in the BCJL protocol,     
where the density matrices $\rho_0$ and $\rho_1$ are      
not identical. 
We show that if the protocol is secure against Bob, then it
is not secure against Alice.  We must start with a necessary
and natural criteria for the security against Bob.  We use a
criteria that makes sense for anyone who understands what it
means to guess the value of a secret bit.  
Let $X$ be the random variable which represents 
the best guess for the bit $b$ chosen by Alice
that can be made by Bob after the commit phase.
Let ${\bf b} = b$ if and only if the bit chosen by Alice is $b$.   
The probability of error for this guess is 
$
PE =  \sum_{b = 0}^1 \Pr({\bf b} = b) \Pr(X = \bar{b} | {\bf b} = b ).
$
Now, let $\epsilon = 1/25$.  The criteria is that, for a
perfectly random bit ${\bf b}$ chosen
by Alice, we must have $| PE - \frac{1}{2} | \leq \epsilon$.
A probability of error close to $1/2$ is is a natural criteria 
to indicate that one did not gain much information 
about a bit ${\bf b}$ that is initially perfectly random.  
We~denote $X_b$ the random variable $X$ conditioned by ${\bf b} = b$     
so that \mbox{$\Pr(X = x\; |\; {\bf b} = b) = \Pr(X_b = x)$}.       
The Kolmogorov distance $K(p_0,p_1)$ 
between two distributions of probability $p_0$,
$p_1$ on a set $A$ is defined by
$
K(p_0,p_1) = \sum_{x\in A} |p_0(x) - p_1(x)|. 
$
Let $p_b(x) = \Pr(X_b = x)$. 
After some algebra, one obtains that the criteria 
$| PE - \frac{1}{2} | \leq \epsilon$ implies that
$K(p_0,p_1) \leq 4 \epsilon$.  
However, this inequality has been obtained for values
of $K$ that are defined in terms of 
measurements that return two outcomes, 
whereas the Kolmogorov distance $K$ can be defined
for an arbitrary number of outcomes. Let us  show that, if the inequality 
$K \leq 4 \epsilon$ holds for any binary outcome measurement, 
the same inequality holds for an arbitrary measurement. 
It is shown  in \cite{davies76,mayers96d} that 
the most general measurement on the $n$ photons that is
allowed by quantum mechanics can be 
described by operators $M_1^{(B)}, \ldots , M_k^{(B)}$ 
given by equations $M_j^{(B)} = P_j U$ 
where $U$ is an isometry from the space of
the $n$ photons to some other Hilbert space $H$ and the operators
$P_j$ are projection operators that define an orthogonal
measurement on $H$.  The exponent $(B)$ reminds us
that the measurement is executed by Bob.  
The classical outcome $j$ returned by
this measurement is the value taken by a random variable $J$.
Again, we denote $J_b$ the random variable     
$J$ conditioned by ${\bf b}  = b$.     
Let $A$ be the set of possible values for $J$.  
Let $A_0 = \{ j \in  A \; | \;     
\Pr(J_0  = j ) \geq \Pr(J_1 = j ) \}$ and $A_1 = A - A_0     
= \{ j \in A \; | \; \Pr(J_1 = j ) > \Pr(J_0 = j)\}$.     
We define $M'_0 = \sum_{j \in A_0} M_j$ and     
$M'_1 = \sum_{j \in A_1} M_j$.  One may easily check that      
$M'_0$ and $M'_1$ define an incomplete measurement with a binary     
classical outcome. Let  $X$ be the random binary outcome of this 
measurement. We have that  $\Pr(X_b = x ) = \Pr(J_b \in A_x )$.  
As desired we obtain: 
\begin{eqnarray*}     
K &= & \sum_{j \in A} | \Pr(J_0 = j) - \Pr(J_1 = j) | \\     
  &= & \sum_{j \in A_0} \Pr(J_0 = j) - \Pr(J_1 = j)    \\
& &  + \; \sum_{j \in A_1} \Pr(J_1 = j) - \Pr(J_0 = j)  \\    
  &= & ( \Pr(J_0 \in A_0 ) - \Pr(J_1 \in A_0) ) \\    
& &  + \; ( \Pr(J_1 \in A_1 ) - \Pr(J_0 \in A_1 ))\\     
& = & |  (\Pr(X_0 = 0 ) - \Pr(X_1 = 0 ) )| \\
& &  +\; | (\Pr(X_1 = 1) - \Pr(X_0 = 1 ))|\\
& \leq & 4 \epsilon       
\end{eqnarray*}     
Now, let us consider the Bhattacharyya-Wootters 
distance~\cite{fuchs95b,toussaint71c,wootters81}
\[     
BW = \sum_{j\in A} \Pr(J_0 = j)^{\frac{1}{2}} \Pr(J_1 = j)^{\frac{1}{2}}.
\]     
It is explained in \cite{toussaint71c} that $(1 - BW) \leq K/2$.      
Therefore, we have $BW \geq (1 - 2 \epsilon)$.        
Furthermore, in \cite{fuchs95b,wootters81} it is shown that the minimum     
of $BW$ over all possible measurement is      
the fidelity $F$ between $\rho_0$ and $\rho_1$.     
So, we have $1 \geq F \geq  (1 - 2 \epsilon)$.       
A purification of $\rho_b$ is simply a pure state
of the overall system that has $\rho_b$  for density
matrix on Bob's side.  
A theorem due to Uhlmann \cite{jozsa94a,uhlmann76} 
says that the fidelity between two     
mixed states $\rho_0$ and $\rho_1$ is given by     
\[     
F = \max |\langle \phi_0 | \phi_1 \rangle|^2     
\]     
where the maximum is taken over the purifications $\phi_0$     
and $\phi_1$ of $\rho_0$ and $\rho_1$ respectively.
Therefore, there exists two purifications
$\phi_0$ and $\phi_1$ such that      
\[     
\langle \phi_0 | \phi_1 \rangle^2 = F \geq 
(1 - 2 \epsilon). 
\]     
We describe Alice's  strategy.     
Alice prepares the incremented system in the state     
$|\phi_0\rangle$.  Clearly, 
if Alice prepares the state $|\phi_0\rangle$, 
she can choose a measurement     
${\bf M}_0^{(A)}$ that returns $\psi = (\theta,c)$ which     
will convince  Bob that she had $b = 0$ in mind.     

Now, assume that Alice has prepared $|\phi \rangle
= |\phi_0\rangle$, but wants to convince Bob that     
she had $b =1$ in mind. 
We show that the measurement  ${\bf M}^{(A)}_1$
that works when Alice prepares the state $|\phi_1\rangle$, 
works as well even if Alice has prepared the state $|\phi_0\rangle$.     
On Bob's side we may consider that Bob executes
a measurement ${\bf M}^{(B)}$ 
that computes  $(\hat{\theta},\hat{c})$.
Alice's measurement ${\bf M}^{(A)}_1$ 
and Bob's measurement ${\bf M}^{(B)}$, both together,     
determine an overall measurement ${\bf M}_1$      
on the overall system.  
The classical outcome of this overall measurement 
is denoted $y = (\theta, c, \hat{\theta}, \hat{c})$.
This measurement is determined by an isometry $U_1$
and projection operators $P_{1,y}$ that define an
orthogonal measurement on the image of $U_1$~\cite{davies76,mayers96d}.  
We have that $M_{1,y} = P_{1,y} U_1$ is the collapse 
operator associated with the outcome $y$.
We have $\Pr(X = y \; |\; \Phi =  \phi) 
= \|M_{1,y} | \phi\rangle \|^2$ and   
$\Pr(T_1(Y) = ok\; | \; \Phi = \phi) 
= \|M_{1,ok}| \phi\rangle \|^2$,
where $M_{1,ok} = \sum_{y\; ;\; T_1(y) = ok } M_{1,y}$.
We obtain:
\begin{eqnarray*} 
\lefteqn{ | \Pr(T_1(Y) = ok \; | \; \Phi = \phi_0 )} \\
\lefteqn{ - \Pr(T_1(Y) = ok\; | \; \Phi = \phi_1 ) | } \\
& & = |\; \| M_{1,ok}| \phi_0\rangle\|^2 - \|M_{1,ok} |\phi_1\rangle\|^2 \; |\\
& & \leq 2 \times  \| M_{1,ok} (|\phi_0\rangle - \phi_1\rangle) \| \\
& & \leq 2 \times \| \;( |\phi_0\rangle - |\phi_1\rangle)\; \| \\
& & =  2 \times \sqrt{2 (1 - \langle \phi_0|\phi_1\rangle) } 
   \leq 4 \sqrt{\epsilon}.
\end{eqnarray*}
If the protocol is correct, we can also assume that 
$\Pr(T_1(Y) = ok\; | \; \Phi = \phi_1 ) \geq 1 - \epsilon'$
where $\epsilon' = 1/25$.
Therefore, we obtain that
$\Pr(T_1(Y) = ok\; | \; \Phi = \phi_0 ) \geq 1 - \epsilon' - 4 \sqrt{\epsilon}
= 4/25$.  This concludes the proof that the BCJL protocol  is insecure.

\paragraph*{Aknowledgment}
The author gratefully acknowledge fruitful discussions with Gilles
Brassard, Claude Cr\'epeau, Christopher Fuchs, Jeroen van de Graaf
Louis Salvail and Williams Wootters.  The author is particularly thankful
to Williams Wootters to have explained 
how to address the simple case while he was in Montr\'eal during
the winter 1995 semester, to Louis Salvail because by claiming  that 
the  simple case does not applied to the BCJL protocol he
actually  raised the important question and
to Jeroen van de Graaf to have pointed out
a result of Christopher Fuchs and a result of Toussaint 
which have been used 
to address the case that really applies to the BCJL protocol.

\end{document}